\documentclass[10pt,prl,twocolumn,superscriptaddress,preprintnumbers,amsmath,amssymb,floatfix]{revtex4-2}

\usepackage{amsmath}
\usepackage{mathrsfs}
\usepackage{soul}

\DeclareFontFamily{U}{calligra}{}
\DeclareFontShape{U}{calligra}{m}{n}{<->callig15}{}

\newcommand{\calE}{{\!\!\text{\usefont{U}{calligra}{m}{n}E}\,\,}}
\newcommand{\calB}{{\!\!\text{\usefont{U}{calligra}{m}{n}B}\,\,}}

\usepackage[T1]{fontenc}
\usepackage{currvita}

\usepackage{graphics,amssymb,amsmath,epsfig,color}
\usepackage{graphicx}
\usepackage{bm}

\usepackage{dcolumn}
\usepackage{bm}
\usepackage{float}
\usepackage{natbib}

\usepackage[export]{adjustbox}

\usepackage{braket}
\setcitestyle{square}

\usepackage{graphicx}
\usepackage{dcolumn}
\usepackage{bm}
\usepackage{float}
\usepackage{natbib}
\setcitestyle{square}

\usepackage{xcolor}

\usepackage{hyperref}
\hypersetup{colorlinks=true, linkcolor=blue, citecolor=blue, urlcolor=blue}

\begin{document}

\title{Handedness selection and hysteresis of chiral orders in crystals}

\author{Mauro Fava}
 \affiliation{Physique Th\'eorique des Mat\'eriaux, QMAT, Universit\'e de Li\`ege, B-4000 Sart-Tilman, Belgium}

\author{Aldo H. Romero}
\affiliation{Department of Physics and Astronomy, West Virginia University, Morgantown, WV 26505-6315, USA}

\author{Eric Bousquet}
\affiliation{Physique Th\'eorique des Mat\'eriaux, QMAT, Universit\'e de Li\`ege, B-4000 Sart-Tilman, Belgium}

\date{\today}

\begin{abstract}
\noindent A phase transition can drive the spontaneous emergence of chiral orders in crystals below a critical temperature.
However, selecting either a right- or a left-handed phase with the aid of electromagnetic fields is challenging, particularly when intrinsic polar and axial moments are lacking. 
In this work we show that \textit{purely} chiral phases with opposite handedness, when both deriving from one degenerate instability, are linked by accessible transition states.
While these states compete with the chirality under an electromagnetic field, a circularly polarized source can select the handedness of the system. This selection is mediated by a chiral monopole and may further result in a hysteresis process of the gyrotropic properties, namely the optical activity, below the critical temperature. 
We suggest several materials, among which 
K$_3$NiO$_2$, as candidates for possible experimental observation.
\end{abstract}

\maketitle

\textit{Introduction.--} Many systems in nature are chiral and cannot be superimposed on their mirror images. 
While the structural chirality is often fixed by the crystallization process, it can emerge as a time-reversal even order parameter~\cite{hlinka2014,chiroaxial_Erb_Hlinka} (as an active degree of freedom) and also spontaneously appear below a finite critical temperature (T$_c$)~\cite{Kimura2016,PhysRevB.102.235127,Hayashida2021,Hayashida2022,PhysRevB.108.L201112,PhysRevB.109.024113}. 
Following ref.~\cite{PhysRevLett.129.116401}, 
a chiral monopole G$_0$, a time-reversal even polarization $\mathbf{P}$ and an electric toroidal (axial, parity even) $\mathbf{G}$ order~\cite{DUBOVIK1990145,PhysRevLett.96.237601,
PhysRevLett.116.177602,PhysRevLett.130.256801} couple via the $G_0 \cdot \mathbf{P}\cdot \mathbf{G}$ energy invariant. 
This suggests that a handedness (right, R and left, L) 
of choice could be selected in some crystals with the help of electromagnetic (EM) fields that control the sign of $\mathbf{P}\cdot\mathbf{G}$. 
In particular, a linearly polarized electric field may be used if either $\mathbf{P}$ or $\mathbf{G}$ are intrinsically present~\cite{Khalyavin,Romao2024},
as experimentally realized~\cite{doi:10.1126/sciadv.abj8030}
in the case of spontaneous helical dipolar textures and photo-induced chiral distortions~\cite{doi:10.1126/science.adr4713}.
In particular, helical textures show a hysteresis of the handedness (R $\leftrightarrow$ L \textit{conversion} below T$_c$) 
under a DC field~\cite{doi:10.1126/sciadv.abj8030}.
However, chiral materials possess neither polar nor axial orders in general.
In this case, a circularly polarized EM field is necessary to select a handedness~\cite{Xu2020}, and heating-cooling cycles with respect to T$_c$ seem to be required to convert it (R $\leftrightarrow$ L) when a spontaneous chiral order is present~\cite{Xu2020,Hayashida2022}, possibly due to a lack of a suitable transition path.
It is hence questionable whether the handedness of a spontaneous chiral order below T$_c$ can generally be controlled.

With the help of density functional theory (DFT) and Landau theory, we show that a way to overcome this hindrance is to consider a degenerate chiral instability that induces both right- and left-handed phases, so that 
a low-energy transition path connecting R and L is guaranteed. 
Therefore, we analyze the nonlinear interaction between the chirality and induced polar and electric toroidal moments as a way to realize handedness conversion. 
In particular, we distinguish between selective and non-selective effects that are simultaneously activated by an external source.
First, we show that a field-mediated competition between chiral and transition states can take place. The latter are particularly favored if the external source is linearly polarized, which hinders the conversion process. 
Nevertheless, through static and dynamical calculations, we show that selecting a handedness and realizing a hysteresis of the chirality and associated properties (namely the natural optical activity) below T$_c$ is feasible if a $\mathbf{P}\cdot\mathbf{G}$ monopole is switched on, for instance by a chiral electromagnetic field. 
While evaluating our model on the K$_3$NiO$_2$ compound~\cite{djurivs2012k3nio2},
this mechanism applies to several families of materials, as long as one degenerate instability generates both right- and left-handed chiral phases.

\begin{figure}
     \includegraphics[width=0.45\textwidth]{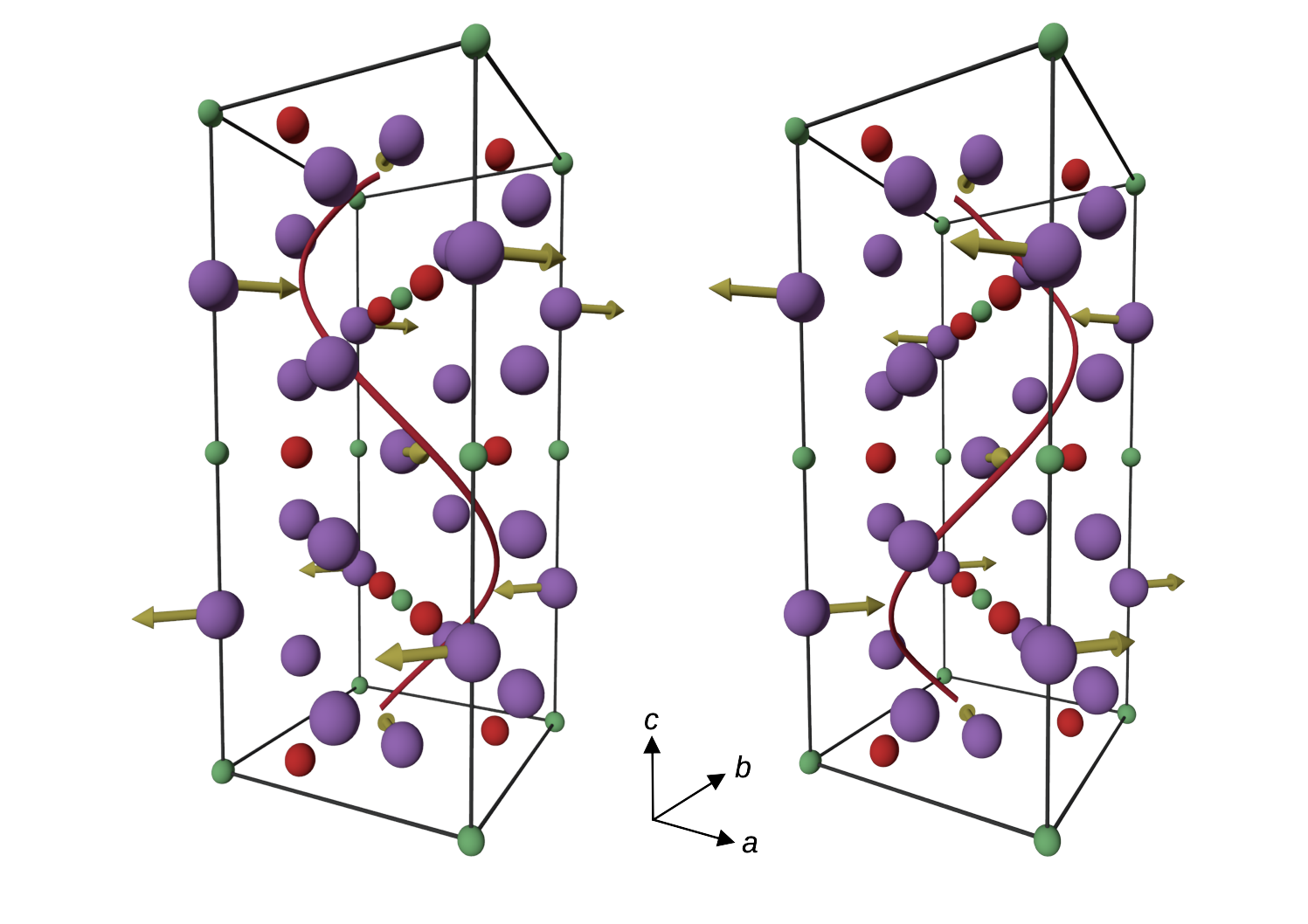}
     \caption{Scheme of the nonequivalent handed spirals spontaneously induced by a degenerate instability in K$_{3}$NiO$_{2}$. The purple, green, and red spheres correspond to K, Ni, and O atoms. The ions are in the $P4_{2}/mnm$ configuration (1x1x2 supercell). At the same time, the golden arrows and the red spirals represent the $Z_{4}$ chiral ($P4_{1}2_12$ left, $P4_32_12$ right) displacement (for clarity, only the main K-motion is shown).}
     \label{fig:figure_1}
     \end{figure}

\begin{figure}
     \includegraphics[width=0.45\textwidth]{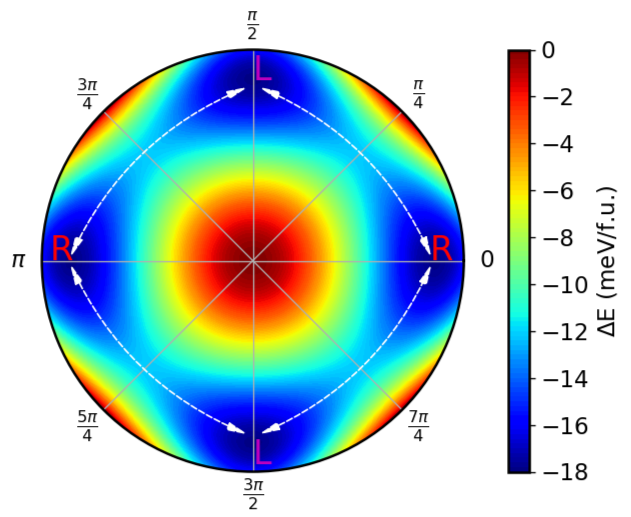}
     \caption{Energy surface F = $\alpha\phi^2+\{\beta_1+\beta_2\sin^2{2\theta}\}\phi^4$ of the Z mode instability of K$_3$NiO$_2$ 
     (amplitude $\phi$ in arbitrary units). 
     The origin corresponds to the high symmetry $P4_2/mnm$ structure.
     The chiral $P4_12_12$ and $P4_32_12$ phases are located at $\theta=\{0,\pi\}$ and $\theta=\{\pi/2,3\pi/2\}$ respectively, while the $Cmcm$ phases is at $\theta=\{\pi/4+k\pi/2\}$. 
     The white arrows indicate possible transition paths connecting R and L through low-energy $Cmcm$ transition states.}
     \label{fig:energy_landscape}
     \end{figure}

\textit{Chiral degenerate instability and transition paths.--} 
A degenerate instability is not a fundamental requirement for the spontaneous emergence of a chiral order; nevertheless, such an occurrence is not uncommon in nature.
For instance, it can be expected if the resulting right- and left-handed low-symmetry phases are \textit{enantiomorphs}, namely their respective space groups are different, as is the case for many materials such as K$_3$NiO$_2$~\cite{djurivs2012k3nio2}, 
$ABX_3$ halide perovskites (e.g. CsCuCl$_3$)~\cite{Yamamoto2021,hirotsu1977}),
$AB_2X_4$ spinels (e.g. MgTi$_2$O$_4$)~\cite{Isobe2002,Schmidt2004,PRL_orbital_order,PRL_Peierls} or SiO$_2$ cristobalite~\cite{PhysRevB.46.1,V_Dmitriev_1997,Leadbetter,Hatch1991}. 
For symmetry reasons, the triggering chiral distortions $\ket{R}$ and $\ket{L}$ must be independent.
As follows, we consider an insulator, K$_3$NiO$_2$ (KNO), as a practical example. 
The experimental high symmetry phase of KNO adopts an achiral tetragonal structure with $P4_{2}/mnm$ (no. 136) space group, and a first-order phase transition to low-temperature phases of either right-handed $P4_{1}$2$_{1}2$ (no. 92) or left-handed $P4_{3}$2$_{1}2$ (no. 96) symmetries has been measured on cooling below $\sim$ 423 K~\cite{djurivs2012k3nio2}. 
A symmetry mode analysis~\cite{isodistort} 
of the experimental structures shows that the transition is induced and chiefly described by an instability at the Z = (0,0,1/2) point of the Brillouin zone, with the Z$_4$ doubly degenerate representation. 
This analysis is supported by our DFT + U calculations~\cite{supp,Kresse1999,Perdew1996,PhysRevB.57.1505,PhysRevB.71.035105} (U = 4.2 eV~\cite{PhysRevB.71.035105}), highlighted in a separate work~\cite{favaKNO},
of the phonons~\cite{phonopy1,phonopy2} and relaxed structures. 
Indeed, we identify a doubly degenerate unstable mode at the Z point, with frequency $\omega_\text{Z}$ = 1.35i THz and Z$_4$ irreducible representation that leads to the R and L low-symmetry ground states (see Fig.~\ref{fig:figure_1}). 
As a consequence of the degeneracy, arbitrary $\ket{\psi(\theta)}\equiv\sin{\theta}\ket{R}+\cos{\theta}\ket{L}$ combinations of right- and left-handed eigenstates can be considered, where $\theta$ is a mixing angle.
Thus, the Landau free energy expression F reads~\cite{favaKNO,INVARIANTS}:

\begin{equation}\label{eq:free_energy}
    F(\phi,\theta) = \alpha\phi^2+\{\beta_1+\beta_2\sin^{2}{2\theta}\}\phi^4,
\end{equation}

where $\phi$ is the amplitude of the distortion. While choosing $\theta=\{k\pi,k\pi+\pi/2\}$ corresponds to the $\ket{R}$ and $\ket{L}$ distortions  respectively, their $\ket{R}\pm\ket{L}$ combinations possess $Cmcm$ (no. 63) non-chiral symmetry. 
A generic $\theta\neq\{\pi/4+k\pi/2,\pi/2+k\pi,k\pi\}$ lowers the symmetry to $C222_1$ (no. 20) instead. 
The $\beta_2$ parameter breaks the rotational invariance of the energy surface in Eq.~\ref{eq:free_energy}. 
As such, and upon full structural relaxation 
we compute a $\Delta$E = 17.9 meV per formula unit 
(f.u.) energy gain for both the R and L ground states. 
This gain reduces to $\Delta$E = 12.2 meV f.u. 
in the (degenerate) $Cmcm$ phases, denoting possible competition with the chiral orders. 
Importantly, epitaxial constraints and hydrostatic pressure may tune these energy scales~\cite{favaKNO}. 
Also, Eq.~\ref{eq:free_energy} is additionally justified since secondary 
distortions are negligible~\cite{favaKNO}. 
A free energy as in Eq.~\ref{eq:free_energy} is universal whenever $\ket{R}$ and $\ket{L}$ span a multi-dimensional space~\cite{INVARIANTS}. 
This suggests low-energy L $\leftrightarrow$ X $\leftrightarrow$ R transition paths (absent if the instability is non-degenerate) to convert the handedness of a crystal below T$_c$, where the $X$ orthorhombic transition state
is energetically closer to the chiral structures compared to the reference configuration. In Fig.~\ref{fig:energy_landscape} we show these paths in the case of KNO (X = $Cmcm$ phases).

\begin{figure*}[htb!]
     \includegraphics[width=0.7\textwidth]{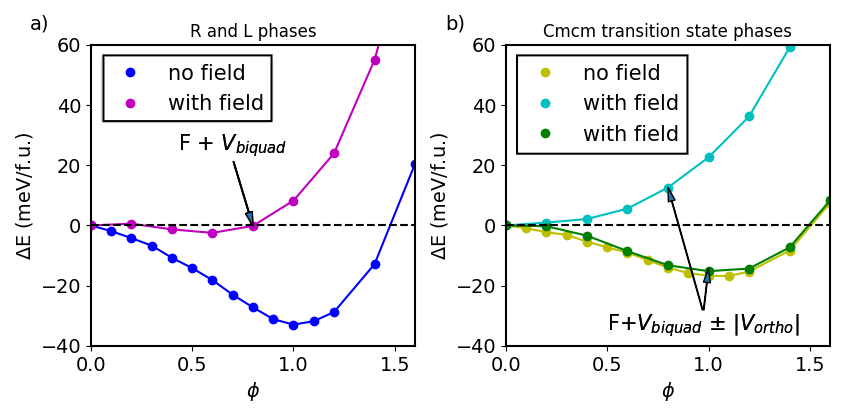}
     \caption{Density functional theory energy landscape as a function of the $\phi$ amplitude (arbitrary units), with/without an electric field [1,1,1] x 
     4.63x10$^{-2}$ V/nm     
     and no axial conjugate field. F, V$_\text{biquad}$ and V$_\text{ortho}$ are defined in the main text. The induced chiral-polar-axial interaction cannot generate any R-L splitting but allows a) an up-shift of the R and L energies and b) the up-shift and splitting ($\pm$V$_\text{ortho}$) of the transition states $Cmcm$ energies.} 
     \label{fig:w_field_no_strain}
     \end{figure*}

\textit{Chiral order / EM field interaction.--}
To see if an external electromagnetic field can be used to 
trigger a permanent handedness conversion, we need to understand how the energy surface interacts with parity-odd polar $\mathbf{P}$ ($\Gamma_{3}^{-}\oplus\Gamma_{5}^{-}$ representation) and parity-even axial $\mathbf{G}$ ($\Gamma_{3}^{+}\oplus\Gamma_{5}^{+}$ representation) distortions generated by such field.
At the lowest (fourth) order, this interaction is given by:

\begin{equation}\label{eq:chiral_polar_axial_interaction}
\begin{split}
        V^{(4)} = \phi^2\sum_{i=x,y,z}(a_iP_i^2+b_iG_i^2) + \\ 
        + \phi^2\cos{2\theta}\sum_{i=x,y,z}\lambda_iP_iG_i + \\
        + \phi^2\sin{2\theta}[\gamma_1P_xP_y + \gamma_2G_xG_y],
\end{split}
\end{equation}

and we can evaluate its effect on the energy surface as follows. The isotropic $V_\text{biquad}\equiv\phi^2\sum_{i}(a_iP_i^2+b_iG_i^2)$ term does not break any symmetry of Eq.~\ref{eq:free_energy}. The second 
term $V_\text{flip}\equiv\phi^2\cos{2\theta}\sum_{i}\lambda_iP_iG_i$ instead crucially lifts the degeneracy of the right- and left-handed phases by coupling the intrinsic chiral order to $P_iG_i$ monopoles.  
The final $V_\text{ortho}\equiv\phi^2\sin{2\theta}[\gamma_{1}P_xP_y+\gamma_{2}G_xG_y]$
term does not affect the R and L structures, but splits the energies of nonequivalent transition states (generated by $\ket{R}+\ket{L}$ and $\ket{R}-\ket{L}$ respectively). 
We show in the supplementary section II~\cite{supp} that an interaction such as Eq.~\ref{eq:chiral_polar_axial_interaction} and a 2D chiral order parameter are present in various different materials, thus suggesting a common phenomenology for handedness selection.
In absence of intrinsic polar and electric-toroidal moments, a finite $\mathbf{P}\cdot{\mathbf{G}}$ monopole \textit{must} be induced to generate a net chiral effect in the material~\cite{PhysRevLett.129.116401}. 
While a static linearly polarized electric field $\calE$ can trigger both polar $\mathbf{P}$ and electric axial distortions, the latter $\mathbf{\tilde{G}}_{xy}\sim(\mathbf{P}_{xy}\times\mathbf{P}_z)$ as a dipolar torque~\cite{PhysRevLett.118.054101,INVARIANTS},
one can easily see that the resulting $\mathbf{P}\cdot\mathbf{\tilde{G}}_{xy}$ monopole is zero and thus no handedness can be selected in tetragonal, hexagonal, and cubic systems (e.g. if $\lambda_x=\lambda_y$ in Eq.~\ref{eq:chiral_polar_axial_interaction}).
On the other hand, a related but different electric toroidal order 
$\mathbf{G}\equiv\sum_{i=1}^{N}\mathbf{u}_{xy,i}\times\mathbf{p}_{zi}/\Omega_\text{u.c.}$~\cite{PhysRevLett.96.237601,PhysRevResearch.6.043141} originates from the multipole expansion of the polarization density field $\mathbf{P}(\mathbf{r})$~\cite{supp,PhysRevB.93.195167}. 
Hence, the interaction between $\mathbf{P}$, $\mathbf{G}$ and an EM field can be expanded as:

\begin{equation}\label{eq_chiral_source}
    V = -\mathbf{P}\cdot\mathbf{\calE} - \mathbf{G}\cdot\mathbf{\nabla}\times\mathbf{\calE} - \text{other multipoles}
\end{equation}

which results in $\mathbf{P}\sim\calE$ and $\mathbf{G}\sim
\mathbf{\nabla}\times\calE$~\cite{PhysRevLett.96.237601,PhysRevResearch.6.043141}. The R-L energy difference is thus given by $\sim\phi^2\sum_{i}\lambda_i\calE_i(\nabla\times\calE)_i$, which couples the dynamical chirality of an electromagnetic field 
$\rho_{\chi} \equiv \frac{\epsilon_{0}}{2}\calE\cdot(\mathbf{\nabla}\times\calE) + \frac{1}{2\mu_{0}}\calB\cdot(\mathbf{\nabla}\times\calB)$ ~\cite{PhysRevLett.104.163901} to that of the material, as experimentally shown in ref.~\cite{Xu2020}.

\textit{Chiral/orthorhombic competition.--} Since an EM source is likely to activate both selective ($V_\text{flip}$) and non-selective effects (V$_\text{biquad}$, V$_\text{ortho}$), it is important to visualize how they compete with each other. 
To support our analysis, we develop a DFT~\cite{gonze_2020,pseudodojo}, static and non-magnetic model isostructural to K$_3$NiO$_2$ by replacing 
$\{\text{K,Ni}\}\rightarrow\{\text{Na,Au}\}$ (see supplementary section III A~\cite{supp} and ref.~\cite{NAO_expt} for the details of its construction). We start with the non-selective processes.

First, we apply a constant linearly polarized electric field~\cite{PhysRevLett.89.117602}
$\calE$ = 4.63x10$^{-2}$ V/nm along the [1,1,1] direction of the high symmetry phase and relax the unit cell. 
Then, we extract the polar and the torque-induced axial modes with polarization $\mathbf{P}_0$ =  (0.25,0.25,0.19) e/nm$^{2}$ 
and electric toroidal moment $\mathbf{G}_0$ = (0.002,-0.002,0.0) e/nm which corresponds to Eq.~\ref{eq_chiral_source} with  $\mathbf{\nabla}\times\calE$ = 0. 
Their effect on the energy landscape is shown in Fig.~\ref{fig:w_field_no_strain}. An up-shift of all energies due to the overall positive isotropic term of Eq.~\ref{eq:chiral_polar_axial_interaction} ($V_\text{biquad}$) 
and the splitting of $Cmcm$ phases ($V_\text{ortho}$) along the $\ket{R}+\ket{L}$ and $\ket{R}-\ket{L}$ directions (cyan and green Fig.~\ref{fig:w_field_no_strain}(b) curves) are observed. 
Moreover, no R-L energy difference is detected, consistently with the symmetries of the unit cell. 
While an additional strain deformation $\epsilon_{xx}$ breaks the tetragonal symmetry and introduces a $\mathbf{P}\cdot\mathbf{G}$ monopole, the calculated R-L splitting $\sim(\lambda_x-\lambda_y)\phi^2P_xP_yP_z$
is found to be negligible (supplementary section III B~\cite{supp}). 
Clearly and as shown in Fig.~\ref{fig:w_field_no_strain}, a $Cmcm$ phase may be stabilized by a linearly polarized electric field (temporarily if $\calE\rightarrow\calE e^{-t/\tau}$), which prevents or strongly hinders the selection or conversion of a handedness unless a chiral source is used.

\begin{figure*}[htb!]
\centering
\includegraphics[width=0.7\textwidth]{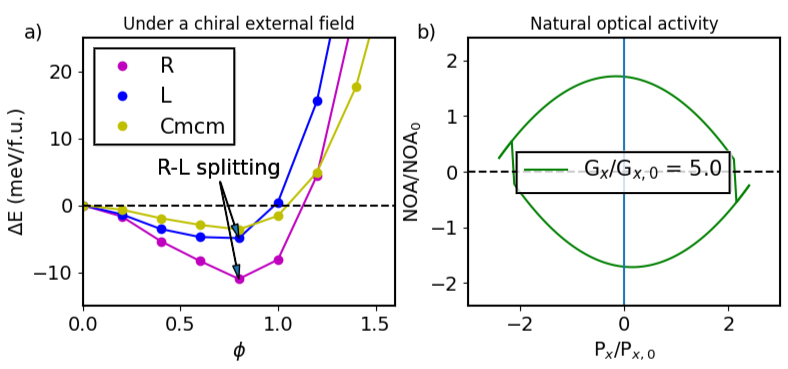}
     \caption{(a) Density functional theory energy landscape as a function of the $\phi$ amplitude (arbitrary units) and in the presence of independent induced polar and toroidal moments along the $x$ axis. $P_{x,0}$ and $G_{x,0}$ are the reference polar and electric toroidal orders, and we fix $P_x/P_{x,0}$ = 1 and $G_x/G_{x,0}$ = 5 respectively.
     In this setting, the $\ket{R}\pm\ket{L}$ ($Cmcm$) degeneracy is unbroken. We observe the R-L splitting induced by 
     V$_\text{flip}\equiv\lambda_x\phi^2\cos{2\theta}P_xG_x$, along with the onset of a L$\rightarrow$ R crossing via the $Cmcm$ transition state. b) Hysteresis of the natural optical activity ($\phi^2\cos{2\theta}$ normalized by its value at at zero fields), as a function of $P_x/P_{x,0}$. The axial distortion is again fixed at $G_{x}/G_{x,0}$ = 5.}
    \label{figure_4}
     \end{figure*}

\textit{R-L splitting/handedness hysteresis.--}To probe a static net chiral effect, we can simulate a $\mathbf{P}\cdot\mathbf{G}\equiv P_{x,0}G_{x,0}$ monopole by applying the previously calculated polar and axial distortions corresponding to $\mathbf{P}$ $\equiv$ (P$_{x,0}$,0,0) and $\mathbf{G}$ $\equiv$ (G$_{x,0}$,0,0) to the energy surface, which eliminates the splitting of the transition states and downsizes the bi-quadratic terms. We also introduce scaling factors P$_x$/P$_{x,0}$ and G$_x$/G$_{x,0}$ as controllable parameters in our model. This is equivalent to employing a circularly polarized EM source with nonzero chirality~\cite{PhysRevLett.104.163901}, where $\mathbf{\nabla}\times\calE$ is the conjugate field of $G_{x}$.

Our DFT calculations are shown in Fig.~\ref{figure_4}(a), where $P_x$/$P_{x,0}$ and $G_x$/$G_{x,0}$ are chosen to be 1 and 5, respectively. 
Along with global shifts ($V_\text{biquad}$) that preserve the four-well structure of the free energy and clear R-L splitting, we notice that the L and $Cmcm$ phases are now saddle points in the energy landscape, suggesting that the aforementioned L$\rightarrow$ $Cmcm$ $\rightarrow$ R crossing below T$_c$ is possible at rather modest values of P$_{x,0}$ and G$_{x,0}$. 
Clearly, the reciprocal R $\rightarrow$ L transition may be realized by changing the sign of $\mathbf{P}\cdot\mathbf{G}$ = P$_x$G$_x$ (namely, the handedness of the chiral electromagnetic source). 
Thus, the scenario of Fig.~\ref{figure_4}(a), along with favoring the right-handed structure, corresponds to the application of an effective coercive field and to a still experimentally unobserved field-mediated hysteresis of the chiral order handedness below T$_c$, in the absence of intrinsic polar and electric toroidal orders. 
An experimental signature of this hysteresis process would be given by quantities that probe the gyrotropic $\phi^2\cos{(2\theta)}$ term of Eq.~\ref{eq:chiral_polar_axial_interaction}, namely the natural optical activity (NOA) tensor in insulators~\cite{PhysRevLett.131.086902} or the circular photogalvanic current in metals~\cite{Xu2020}. 
In Fig.~\ref{figure_4}(b) we show that the optical activity exhibits a hysteresis behavior as a function of the electric and axial conjugate fields (the parameters of the model fitted from DFT are reported in supplementary section IV A~\cite{supp}). 
Bi-quadratic fluctuations ($\phi^2P_x^2$, $\phi^2G_x^2$) constrain the magnitude of the fields and, if too large, can suppress the four-well structure of the energy surface~\cite{supp}. 
Finally, we note that selecting the handedness of the system by applying a chiral external field from above T$_c$, while quenching the temperature as in Ref.~\cite{Xu2020}, is also compatible with this picture.

\begin{figure}
     \includegraphics[width=0.45\textwidth]{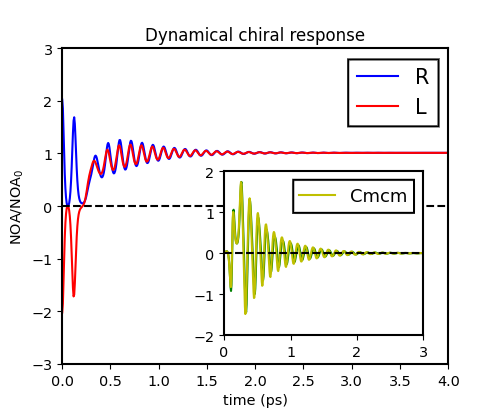}
     \caption{Dynamics of the handedness conversion process. Using a chiral EM plane wave ($\Omega$ = 3.26 THz, $\tau$ = 1/$\omega$, P$_{x}$/P$_{x,0}$ = 1 and G$_x$/G$_{x,0}$ = 0.5) as a source, the ensuing $\mathbf{P}\cdot\mathbf{G}>0$ monopole triggers a stable L$\rightarrow$R conversion over time, while the handedness of the R phase is instead preserved at the end of the evolution, resulting in a net right-handed system. Inset: damped oscillations of the  $\phi^2\sin(2\theta)$ quantity measuring the occupation of the transition states over time.}
     \label{fig:figure_5}
     \end{figure}

\textit{Dynamical effects.--} To better visualize the enantioselectivity process, we consider how the chirality of the system evolves under a circularly polarized short EM pulse with a driving frequency $\omega$ in the THz range (see SI~\cite{supp}, section IV B). 
Treating the generated $\mathbf{P}(t)$ and $\mathbf{G}(t)$ orders as damped harmonic oscillators~\cite{supp,PhysRevLett.96.237601,Romao2024,doi:10.1126/science.adr4713} along the $x$ and $y$ directions, we find that the $\|\mathbf{P}(t)\|$ and $\|\mathbf{G}(t)\|$ amplitudes alongside the resulting $\mathbf{P}(t)\cdot\mathbf{G}(t)$ monopole only depend on time through a $\exp(-t/\tau)$ factor. 
Hence, the chiral-polar-axial interaction is almost equivalent to that of Fig.~\ref{figure_4} in the $\tau\rightarrow\infty$ limit, with the difference that V$_\text{ortho}\propto e^{-t/\tau}\sin(2\omega t)$ is not zero. To gain additional insight, we solve the equations of motion of the system:

\begin{equation}
    M\ddot{\mathbf{\phi}} + \gamma\dot{\mathbf{\phi}} + \mathbf{\nabla}(F+V^{(4)}) = 0, 
\end{equation}

using DFT-obtained parameters. In Fig.~\ref{fig:figure_5} under non-resonant conditions, we show that starting from right- and left-handed phases in the absence of fields at t = 0, if $\mathbf{P}\cdot\mathbf{G}>0$ is activated we have L $\rightarrow$ R and R $\rightarrow$ R, with a net right-handed phase stationary at the end of the evolution when the external fields are suppressed. This is consistent with a permanent and stable conversion of the chirality in a hysteresis fashion below T$_c$, and happens despite having a nonzero V$_\text{ortho}$. Inducing $\mathbf{P}\cdot\mathbf{G}<0$ stabilizes instead a left-handed phase~\cite{supp}.
Using the $\phi^2\sin{2\theta}$ quantity to characterize the occupation of the transition states throughout the dynamics, we further confirm that the final equilibrium state is enantiomorphic ($\phi^2\sin(2\theta)=0$ with $\phi\neq0$).

\textit{Application to other materials.} While suggesting K$_3$NiO$_2$ or analogous $A_3$$B$O$_2$ oxides~\cite{djurivs2012syntheses,favaKNO} for the experimental exploration of the handedness hysteresis process, other systems may be evaluated.
As previously discussed (supplementary section II~\cite{supp}), a degenerate instability at the energy surface, generating chirality below a critical temperature, can be found in $ABX_3$ halide perovskites, $AB_2$$X_4$ spinels and SiO$_2$ cristobalite. 
Additionally, several $A(B,B')_2$X$_4$ spinels may adopt cation-ordered structures of $P4_{1,3}22$ symmetry ~\cite{Talanov2014-ik} (e.g. LiZnNbO$_4$~\cite{Marin1994, Liu2019, PhysRevLett.105.075501}, $B/B'$ ratio = 1:1), a process suitable to be influenced by Eq.~\ref{eq:chiral_polar_axial_interaction}
due to the chiral and multi-dimensional nature of the occupational order parameter. 
Consequently, the here-proposed strategy to use low-energy transition paths for R $\leftrightarrow$ L interconversion has a rather general character and could be extended to many materials.

\textit{Conclusions}.
We have reported how mirror-opposite spontaneous chiral orders in crystals are connected by low-energy transition states when generated by one degenerate instability.
Relying on this condition, we have developed a model to understand how a handedness may be selected by external fields and in the absence of intrinsic polar and axial orders.
Importantly, we have shown that competing non-selective effects, for instance triggered by a linearly polarized electric field, could be present. 
Nevertheless, an external electromagnetic source generating a $\mathbf{P}\cdot\mathbf{G}$ monopole can induce a net chirality in the system and may also favor a permanent handedness conversion below the critical temperature that persists once the field is suppressed. We argue that the hysteresis of the natural optical activity is a signature of this phenomenon, as supported by our static (Fig.~\ref{figure_4}(b)) and dynamical (Fig.~\ref{fig:figure_5}) calculations. 
Additionally, the interaction of Eq.~\ref{eq:chiral_polar_axial_interaction} could be explored at the phononic level, that is if $\mathbf{P}$ and $\mathbf{G}$ are replaced by infra-red and Raman active phonon modes~\cite{doi:10.1126/sciadv.abj8030} that share the same symmetries. 
The present work hence suggests new perspectives about controlling the chirality of materials through external fields.

\section*{Acknowledgements}
The authors acknowledge E. McCabe for fruitful discussions.
Computational resources have been provided by the Consortium des \'Equipements de Calcul Intensif (C\'ECI), funded by the Fonds de la Recherche Scientifique (F.R.S.-FNRS) under Grant No. 2.5020.11.
MF \& EB acknowledge FNRS for support and the PDR project CHRYSALID No.40003544. Work at West Virginia University was supported by the U.S. Department of Energy (DOE), Office of Science, Basic Energy Sciences (BES) under Award DE‐SC0021375. This work used Bridges2 and Expanse at the Pittsburgh Supercomputer (Bridges2) and the San Diego Supercomputer Center (Expanse) through allocation DMR140031 from the Advanced Cyberinfrastructure Coordination Ecosystem: Services \& Support (ACCESS) program, which the National Science Foundation supports grants 2138259, 2138286, 2138307, 2137603, and 2138296.

\bibliography{bibliography}

\end{document}